\pdfoutput=1
\documentclass[12pt,a4paper]{article}

\usepackage{ifthen} 
\newboolean{pdflatex}
\setboolean{pdflatex}{true} 

\newboolean{articletitles}
\setboolean{articletitles}{true} 

\newboolean{uprightparticles}
\setboolean{uprightparticles}{false} 

\def\paperauthors{LHCb collaboration} 
\def\paperasciititle{Observation of sizeable omega contribution to chi_c1(3872)->pi+pi-J/psi decays} 
\def\papertitle{Observation of\\ sizeable $\omega$ contribution\\ to $\chi_{c1}(3872)\to\pi^+\pi^-J/\psi$ decays} 
\def\paperkeywords{{High Energy Physics}, {LHCb}} 
\def\papercopyright{\the\year\ CERN for the benefit of the LHCb collaboration} 
\def\paperlicence{CC BY 4.0 licence}
\def\paperlicenceurl{https://creativecommons.org/licenses/by/4.0/}

\usepackage[top=1in, bottom=1.25in, left=1in, right=1in]{geometry}

\usepackage{microtype}
\usepackage{lineno}  
\usepackage{xspace} 
\usepackage{caption} 

\usepackage{graphicx}  
\usepackage{color}
\usepackage{colortbl}
\graphicspath{{./figs/}} 

\usepackage{amsmath} 
\usepackage{amssymb}
\usepackage{amsfonts}
\usepackage{upgreek} 

\newcommand*\patchAmsMathEnvironmentForLineno[1]{%
\expandafter\let\csname old#1\expandafter\endcsname\csname #1\endcsname
\expandafter\let\csname oldend#1\expandafter\endcsname\csname
end#1\endcsname
 \renewenvironment{#1}%
   {\linenomath\csname old#1\endcsname}%
   {\csname oldend#1\endcsname\endlinenomath}%
}
\newcommand*\patchBothAmsMathEnvironmentsForLineno[1]{%
  \patchAmsMathEnvironmentForLineno{#1}%
  \patchAmsMathEnvironmentForLineno{#1*}%
}
\AtBeginDocument{%
\patchBothAmsMathEnvironmentsForLineno{equation}%
\patchBothAmsMathEnvironmentsForLineno{align}%
\patchBothAmsMathEnvironmentsForLineno{flalign}%
\patchBothAmsMathEnvironmentsForLineno{alignat}%
\patchBothAmsMathEnvironmentsForLineno{gather}%
\patchBothAmsMathEnvironmentsForLineno{multline}%
\patchBothAmsMathEnvironmentsForLineno{eqnarray}%
}

\usepackage{hyperxmp}

\usepackage[pdftex,
            pdfauthor={\paperauthors},
            pdftitle={\paperasciititle},
            pdfkeywords={\paperkeywords},
            pdfcopyright={Copyright (C) \papercopyright},
            pdflicenseurl={\paperlicenceurl}]{hyperref}

\usepackage[colorinlistoftodos,textsize=scriptsize]{todonotes}

\usepackage[bottom,flushmargin,hang,multiple]{footmisc}

\usepackage[all]{hypcap} 

\usepackage{xspace} 
\usepackage{upgreek}

\def\lhcb   {\mbox{LHCb}\xspace}

\def\belle  {\mbox{Belle}\xspace}

\def\cdf    {\mbox{CDF}\xspace}

\def\MagUp {\mbox{\em Mag\kern -0.05em Up}\xspace}

\ifthenelse{\boolean{uprightparticles}}%
{

 \def\Ppi         {\ensuremath{\uppi}\xspace}                 
                  
 \def\Prho        {\ensuremath{\uprho}\xspace}

 \def\Pchi        {\ensuremath{\upchi}\xspace}                 
 \def\Ppsi        {\ensuremath{\uppsi}\xspace}

 \def\PDelta      {\ensuremath{\Delta}\xspace}                 
 \def\PXi         {\ensuremath{\Xi}\xspace}                 
 \def\PLambda     {\ensuremath{\Lambda}\xspace}                 
 \def\PSigma      {\ensuremath{\Sigma}\xspace}                 
 \def\POmega      {\ensuremath{\Omega}\xspace}                 
 \def\PUpsilon    {\ensuremath{\Upsilon}\xspace}
 \let\oldPi\Pi
 \def\PPi         {\ensuremath{\oldPi}\xspace}

 \def\PB      {\ensuremath{\mathrm{B}}\xspace}                 
                  
 \def\PD      {\ensuremath{\mathrm{D}}\xspace}

 \def\PJ      {\ensuremath{\mathrm{J}}\xspace}                 
 \def\PK      {\ensuremath{\mathrm{K}}\xspace}

 \def\Pb      {\ensuremath{\mathrm{b}}\xspace}                 
 \def\Pc      {\ensuremath{\mathrm{c}}\xspace}

 \def\Pi      {\ensuremath{\mathrm{i}}\xspace}

 \def\Ps      {\ensuremath{\mathrm{s}}\xspace}

 \def\thebaroffset{0.0em}
}
{

 \def\Ppi         {\ensuremath{\pi}\xspace}                 
                  
 \def\Prho        {\ensuremath{\rho}\xspace}

 \def\Pchi        {\ensuremath{\chi}\xspace}                 
 \def\Ppsi        {\ensuremath{\psi}\xspace}                 
                  
 \mathchardef\PDelta="7101
 \mathchardef\PXi="7104
 \mathchardef\PLambda="7103
 \mathchardef\PSigma="7106
 \mathchardef\POmega="710A
 \mathchardef\PUpsilon="7107
 \mathchardef\PPi="7105
                  
 \def\PB      {\ensuremath{B}\xspace}                 
                  
 \def\PD      {\ensuremath{D}\xspace}

 \def\PJ      {\ensuremath{J}\xspace}                 
 \def\PK      {\ensuremath{K}\xspace}

 \def\Pb      {\ensuremath{b}\xspace}                 
 \def\Pc      {\ensuremath{c}\xspace}

 \def\Pi      {\ensuremath{i}\xspace}

 \def\Ps      {\ensuremath{s}\xspace}

 \def\thebaroffset{0.18em}
}
\newcommand{\offsetoverline}[2][\thebaroffset]{\kern #1\overline{\kern -#1 #2}}%

\makeatletter
\ifcase \@ptsize \relax
  \newcommand{\miniscule}{\@setfontsize\miniscule{4}{5}}
\or
  \newcommand{\miniscule}{\@setfontsize\miniscule{5}{6}}
\or
  \newcommand{\miniscule}{\@setfontsize\miniscule{5}{6}}
\fi
\makeatother

\DeclareRobustCommand{\optbar}[1]{\shortstack{{\miniscule (\rule[.5ex]{1.25em}{.18mm})}
  \\ [-.7ex] $#1$}}

\def\squark    {{\ensuremath{\Ps}}\xspace}

\def\cquark    {{\ensuremath{\Pc}}\xspace}
\def\cquarkbar {{\ensuremath{\overline \cquark}}\xspace}
\def\ccbar     {{\ensuremath{\cquark\cquarkbar}}\xspace}
\def\bquark    {{\ensuremath{\Pb}}\xspace}

\def\pion   {{\ensuremath{\Ppi}}\xspace}
\def\piz    {{\ensuremath{\pion^0}}\xspace}
\def\pip    {{\ensuremath{\pion^+}}\xspace}
\def\pim    {{\ensuremath{\pion^-}}\xspace}

\def\rhomeson {{\ensuremath{\Prho}}\xspace}
\def\rhoz     {{\ensuremath{\rhomeson^0}}\xspace}

\def\rhopm    {{\ensuremath{\rhomeson^\pm}}\xspace}

\def\kaon    {{\ensuremath{\PK}}\xspace}

\def\KorKbar {\kern \thebaroffset\optbar{\kern -\thebaroffset \PK}{}\xspace}

\def\Kp      {{\ensuremath{\kaon^+}}\xspace}

\def\Dbar    {{\ensuremath{\offsetoverline{\PD}}}\xspace}
\def\D       {{\ensuremath{\PD}}\xspace}

\def\DorDbar {\kern \thebaroffset\optbar{\kern -\thebaroffset \PD}\xspace}

\def\Dp      {{\ensuremath{\D^+}}\xspace}
\def\Dm      {{\ensuremath{\D^-}}\xspace}

\def\DpDm    {\ensuremath{\Dp {\kern -0.16em \Dm}}\xspace}

\def\Dstarb  {{\ensuremath{\Dbar{}^*}}\xspace}

\def\Dstarzb {{\ensuremath{\Dbar{}^{*0}}}\xspace}

\def\B       {{\ensuremath{\PB}}\xspace}

\def\BorBbar {\kern \thebaroffset\optbar{\kern -\thebaroffset \PB}\xspace}

\def\Bd      {{\ensuremath{\B^0}}\xspace}

\def\BdorBdbar {\kern \thebaroffset\optbar{\kern -\thebaroffset \Bd}\xspace}
\def\Bu      {{\ensuremath{\B^+}}\xspace}

\def\Bs      {{\ensuremath{\B^0_\squark}}\xspace}

\def\BsorBsbar {\kern \thebaroffset\optbar{\kern -\thebaroffset \Bs}\xspace}

\def\jpsi     {{\ensuremath{{\PJ\mskip -3mu/\mskip -2mu\Ppsi}}}\xspace}

\def\Y#1S{\ensuremath{\PUpsilon{(#1S)}}\xspace}

\def\theX     {{\ensuremath{\Pchi_{c1}(3872)}}\xspace}

\def\LorLbar     {\kern \thebaroffset\optbar{\kern -\thebaroffset \PLambda}\xspace}

\def\BR         {\BF}

\newcommand{\decay}[2]{\ensuremath{#1\!\to #2}\xspace} 

\def\to                 {\ensuremath{\rightarrow}\xspace}

\def\AT#1     {\ensuremath{A_{\mathrm{T}}^{#1}}\xspace}           

\def\C#1      {\ensuremath{\mathcal{C}_{#1}}\xspace}                       
\def\Cp#1     {\ensuremath{\mathcal{C}_{#1}^{'}}\xspace}                    
\def\Ceff#1   {\ensuremath{\mathcal{C}_{#1}^{\mathrm{(eff)}}}\xspace}        
\def\Cpeff#1  {\ensuremath{\mathcal{C}_{#1}^{'\mathrm{(eff)}}}\xspace}       
\def\Ope#1    {\ensuremath{\mathcal{O}_{#1}}\xspace}                       
\def\Opep#1   {\ensuremath{\mathcal{O}_{#1}^{'}}\xspace}                    

\newcommand{\aunit}[1]{\ensuremath{\text{\,#1}}}       

\newcommand{\tev}{\aunit{Te\kern -0.1em V}\xspace}
\newcommand{\gev}{\aunit{Ge\kern -0.1em V}\xspace}
\newcommand{\mev}{\aunit{Me\kern -0.1em V}\xspace}
\newcommand{\kev}{\aunit{ke\kern -0.1em V}\xspace}
\newcommand{\ev}{\aunit{e\kern -0.1em V}\xspace}
 
\newcommand{\mevc}{\ensuremath{\aunit{Me\kern -0.1em V\!/}c}\xspace}
\newcommand{\gevc}{\ensuremath{\aunit{Ge\kern -0.1em V\!/}c}\xspace}
\newcommand{\mevcc}{\ensuremath{\aunit{Me\kern -0.1em V\!/}c^2}\xspace}
\newcommand{\gevcc}{\ensuremath{\aunit{Ge\kern -0.1em V\!/}c^2}\xspace}

\def\fb   {\ensuremath{\aunit{fb}}\xspace}
\def\invfb   {\ensuremath{\fb^{-1}}\xspace}

\def\gsim{{~\raise.15em\hbox{$>$}\kern-.85em
          \lower.35em\hbox{$\sim$}~}\xspace}
\def\lsim{{~\raise.15em\hbox{$<$}\kern-.85em
          \lower.35em\hbox{$\sim$}~}\xspace}

\def\PDF {PDF\xspace}

\def\evtgen     {\mbox{\textsc{EvtGen}}\xspace}

\def\geant      {\mbox{\textsc{Geant4}}\xspace}

\def\photos     {\mbox{\textsc{Photos}}\xspace}

\def\pythia     {\mbox{\textsc{Pythia}}\xspace}

\def\tell1  {TELL1\xspace}
\def\ukl1   {UKL1\xspace}

\newcommand{\eg}{\mbox{\itshape e.g.}\xspace}
\newcommand{\ie}{\mbox{\itshape i.e.}\xspace}

\newcommand{\lhcborcid}[1]{\href{https://orcid.org/#1}{\hspace*{0.1em}\raisebox{-0.45ex}{\includegraphics[width=1em]{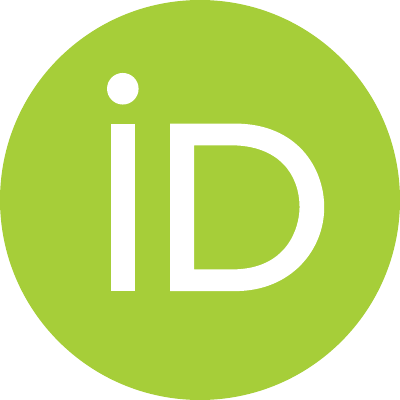}}}}

\usepackage{cite} 
\usepackage{mciteplus}

\usepackage{longtable} 

\begin{document}

\renewcommand{\thefootnote}{\fnsymbol{footnote}}
\setcounter{footnote}{1}


\begin{titlepage}
\pagenumbering{roman}

\vspace*{-1.5cm}
\centerline{\large EUROPEAN ORGANIZATION FOR NUCLEAR RESEARCH (CERN)}
\vspace*{1.5cm}
\noindent
\begin{tabular*}{\linewidth}{lc@{\extracolsep{\fill}}r@{\extracolsep{0pt}}}
\ifthenelse{\boolean{pdflatex}}
{\vspace*{-1.5cm}\mbox{\!\!\!\includegraphics[width=.14\textwidth]{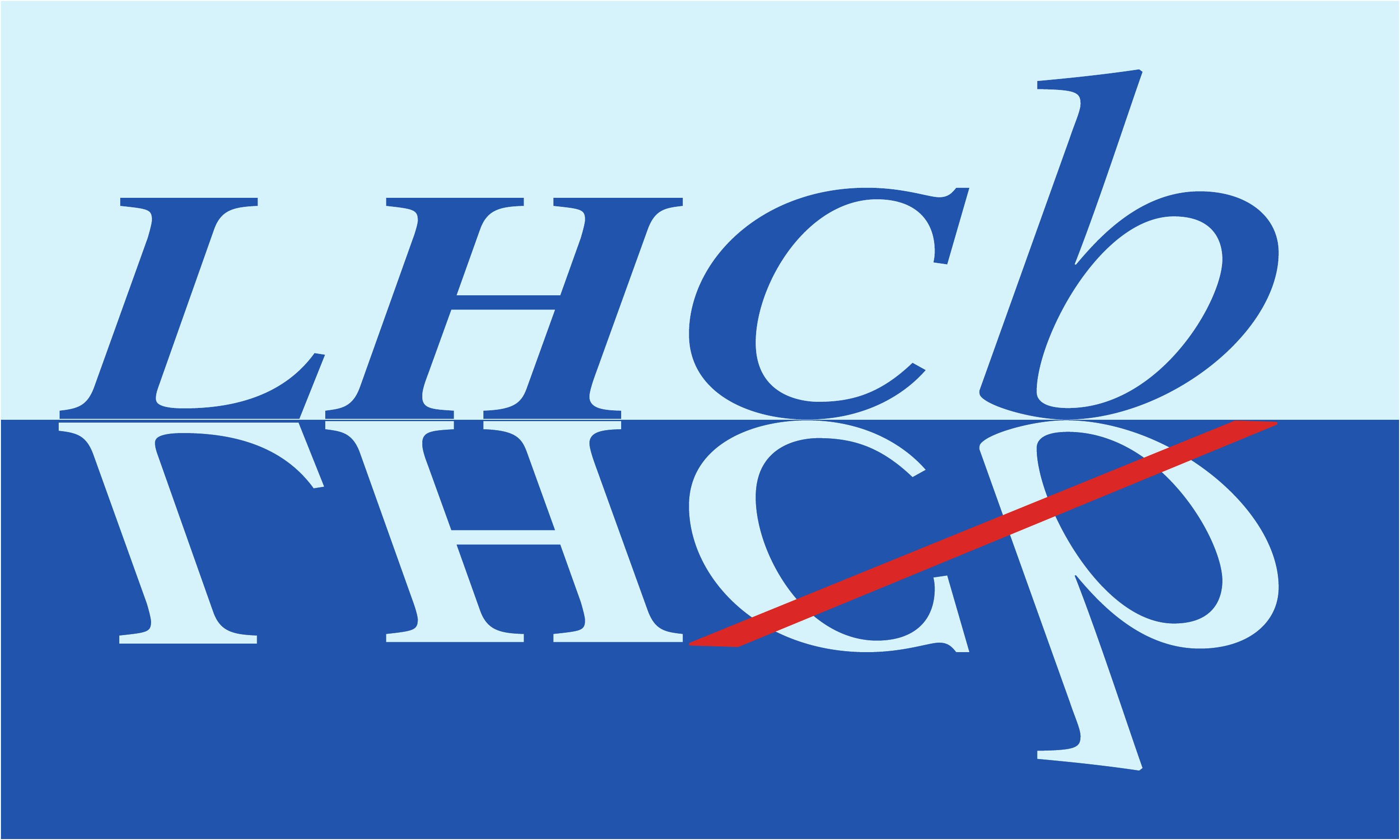}} & &}%
{\vspace*{-1.2cm}\mbox{\!\!\!\includegraphics[width=.12\textwidth]{lhcb-logo.eps}} & &}%
\\
 & & CERN-EP-2022-049 \\  
 & & LHCb-PAPER-2021-045 \\  
 & & August 3, 2023 \\
 & & \\
\end{tabular*}

\vspace*{4.0cm}

{\normalfont\bfseries\boldmath\huge
\begin{center}
  \papertitle 
\end{center}
}

\vspace*{2.0cm}

\begin{center}
\paperauthors
\footnote{Authors are listed at the end of this paper.}
\end{center}

\vspace{\fill}

\begin{abstract}
  \noindent
Resonant structures in the dipion mass spectrum from $\theX\to\pip\pim\jpsi$ decays, produced via $\Bu\to\Kp\theX$ decays,  are analyzed using proton-proton collision data collected by the \lhcb experiment, corresponding to an integrated luminosity of 9\invfb. 
A sizeable contribution from the isospin conserving $\theX\to\omega\jpsi$ decay is established for the first time,  $(21.4\pm2.3\pm2.0)\%$, with a significance of more than $7.1\sigma$.
The amplitude of isospin violating decay, $\theX\to\rhoz\jpsi$, relative to isospin conserving decay, $\theX\to\omega\jpsi$, is  properly determined,
and it is a factor of six larger than expected for a pure charmonium state.

\end{abstract}

\vspace*{2.0cm}

\begin{center}
Published in Phys.~ Rev.~D 131 (2023) L011103 
\end{center}

\vspace{\fill}

{\footnotesize 
\centerline{\copyright~\papercopyright. \href{\paperlicenceurl}{\paperlicence}.}}
\vspace*{2mm}

\end{titlepage}


\newpage
\setcounter{page}{2}
\mbox{~}
%
%
%
%


\renewcommand{\thefootnote}{\arabic{footnote}}
\setcounter{footnote}{0}

\cleardoublepage


\pagestyle{plain} 
\setcounter{page}{1}
\pagenumbering{arabic}


\def\mpp{m_{\pip\pim}}%
\def\mppj{m_{\pip\pim\jpsi}}%
\def\BR{{\cal B}}%
\def\bjkpp{\Bu\to\Kp\pip\pim\jpsi}%
\def\pjpsi{p_{\jpsi}}
\def\PDF{{\cal P}}
\def\M{{\cal M}}
\def\ppi{p_{\pi}}
\def\chindf{\chi^2/{\rm NDoF}}
\def\grpp{g_{\rho\to2\pi}}
\def\gopp{g_{\omega\to2\pi}}
\def\goppp{g_{\omega\to3\pi}}
\def\grppsq{g^2_{\rho\to2\pi}}
\def\goppsq{g^2_{\omega\to2\pi}}
\def\gopppsq{g^2_{\omega\to3\pi}}
\def\Rome{{\cal R}_\omega}
\def\Rall{\Rome^\mathrm{\,all}}
\def\Rzero{\Rome^\mathrm{\,0}}
\def\Riso{{\cal R}_{\omega/\rho}^\mathrm{\,0}}
After the discovery of the $\theX$ state in decays through the $\pip\pim\jpsi$ channel~\cite{Belle:2003nnu}, the $\rhoz\jpsi$ process was suggested to explain the observed $\pip\pim$ mass ($\mpp$) distribution peaking near the upper kinematic limit, close to the $\rhoz$ pole mass. 
The isovector nature of the produced $\pip\pim$ pairs is also supported by the non-observation of $\theX\to\piz\piz\jpsi$ decays.
Since no charged partners of the $\theX$ state have been observed in the $\rhopm\jpsi$ decay mode~\cite{BaBar:2004cah,Belle:2011vlx}, 
the $\theX$ particle is predominantly an isosinglet state;
the $\jpsi$ isospin is also zero. Together, this
makes the $\theX\to\rhoz\jpsi$ decay isospin violating.
The $\theX$ spin and parities, $J^{PC}=1^{++}$~\cite{LHCb-PAPER-2013-001,LHCb-PAPER-2015-015}, match the $2^3P_1$ excitation of the $\ccbar$ system predicted in the relevant mass range~\cite{Eichten:1978tg}. 
However, isospin violating decays of charmonium states are highly suppressed.
Therefore, quantifying the isospin violation in $\theX\to\rhoz\jpsi$ decays is important to understand the nature of the $\theX$ state, which is under intense debate~\cite{Olsen:2017bmm,Guo:2017jvc}. 
Isospin conserving $\theX\to\omega\jpsi$ decays provide a suitable normalization process.
Such decays have been recently established with a significance of $5\sigma$~\cite{BESIII:2019qvy} using the dominant $\omega\to\pip\pim\piz$ decay, which has a branching fraction ($\BR$) of $(89.2\pm0.7)\%$~\cite{PDG2020}.
Averaged with earlier measurements~\cite{Abe:2005ax,delAmoSanchez:2010jr}, 
$\BR(\theX\to\omega\jpsi)/\BR(\theX\to\pip\pim\jpsi)=1.4\pm0.3$.
However, $\omega\to\pip\pim$ decays, with $\BR(\omega\to\pip\pim)=(1.53\pm0.12)\%$~\cite{PDG2020}, are expected to contribute to the denominator at the $2\%$ level if $\rhoz-\omega$ interference is neglected. 
The interference can change this estimate by a large factor. Therefore, an analysis of the $\mpp$ spectrum is necessary to 
disentangle the $\rhoz$ and $\omega$ contributions.
Such analyses were performed by the \cdf~\cite{CDF:2005cfq} and the \belle~\cite{Belle:2011vlx} collaborations, using the Breit--Wigner sum model, yielding inconclusive results due to the large statistical uncertainties.

In this Letter, we report an analysis of a $\Bu\to\Kp\theX$, $\theX\to\pip\pim\jpsi$, $\jpsi\to\mu^+\mu^-$ data sample collected using the \lhcb detector, with proton-proton ($pp$) collision energies of 7, 8 and 13\tev, corresponding to a total integrated luminosity of 9~fb$^{-1}$. The inclusion of charge-conjugate processes is implied throughout. This sample is about six times more sensitive to an $\omega$ contribution than those used in Refs.~\cite{CDF:2005cfq,Belle:2011vlx}. 
The \lhcb
detector~\cite{LHCb-DP-2008-001,LHCb-DP-2014-002} is a single-arm forward spectrometer covering the pseudorapidity range $2 < \eta < 5$, designed for
the study of particles containing \bquark\ or \cquark\ quarks. The detector elements that are particularly
relevant to this analysis are: a silicon-strip vertex detector surrounding the $pp$ interaction
region that allows \bquark\ hadrons to be identified from their characteristically long
flight distance; a tracking system that provides a measurement of the momentum, $p$, of charged
particles; two ring-imaging Cherenkov detectors that are able to discriminate between
different species of charged hadrons 
and the muon detector.

\begin{figure}[tb]
  \begin{center}
    \includegraphics[width=0.7\linewidth]{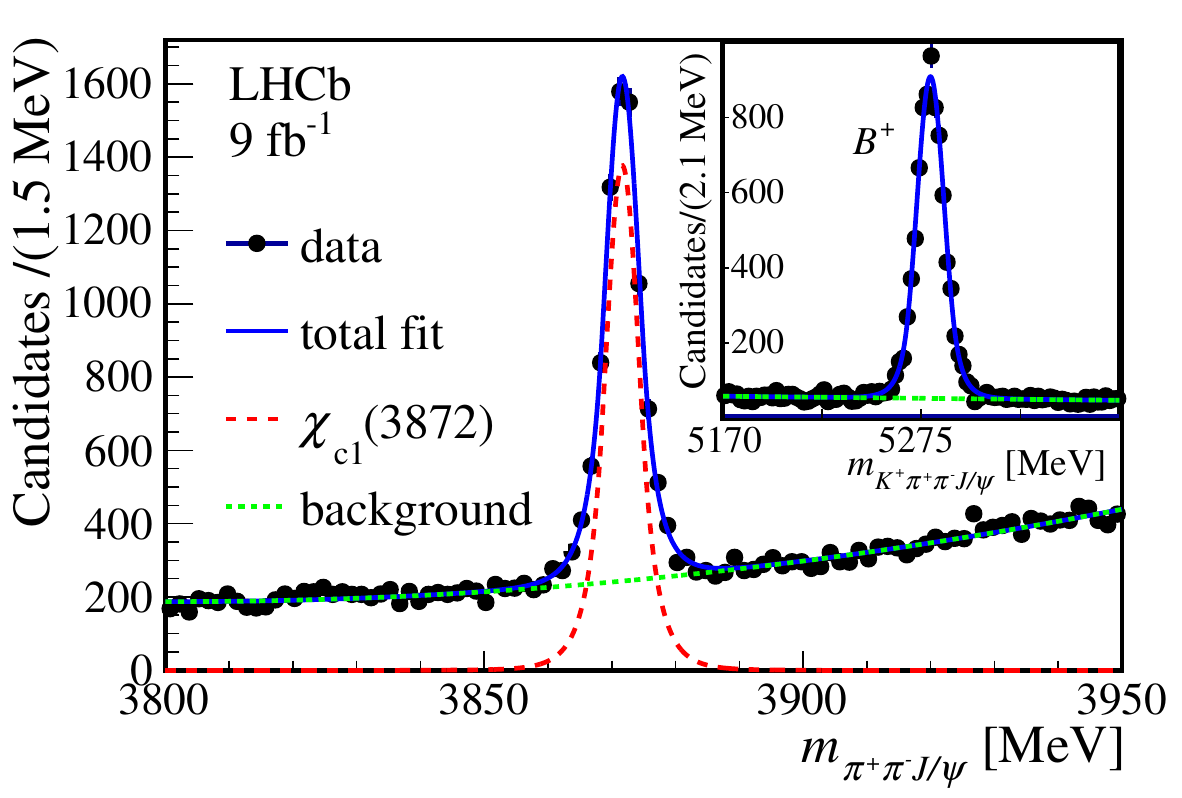}
    \vspace*{-0.5cm}
  \end{center}
  \caption{
    \small 
        Distribution of $\mppj$ ($m_{\Kp\pip\pim\jpsi}$ in the inset) from $\bjkpp$ candidates
        within $\pm2\,\sigma$ around the $\Bu$ 
        ($\theX$) mass, overlaid with the fit including the total (solid blue), signal (dashed red) and background (dashed green) components.
        }
  \label{fig:xmass_fit}
\end{figure}

\begin{figure}[bt]
  \begin{center}
    \includegraphics[width=0.7\linewidth]{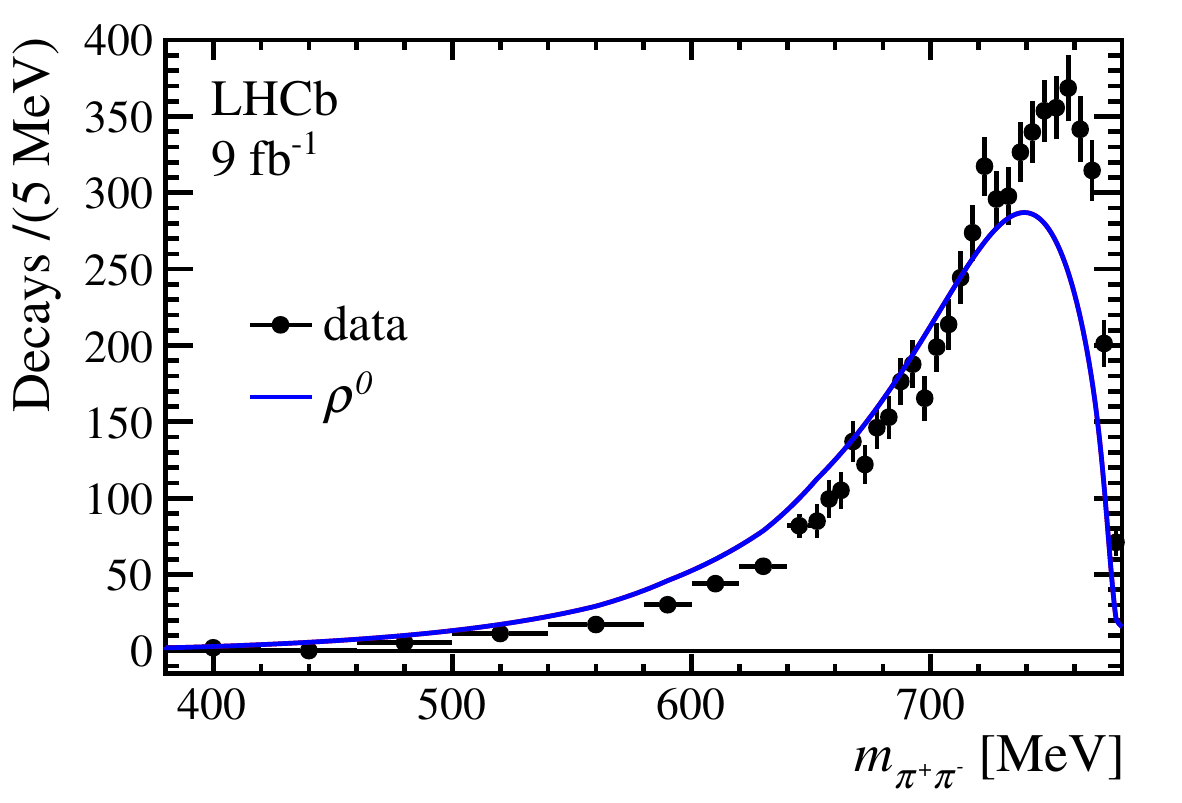}
    \vspace*{-0.5cm}
  \end{center}
  \caption{
    \small 
        Distribution of $\mpp$ in $\theX\to\pip\pim\jpsi$ decays,
        fit with the $\rhoz$-only model.}
  \label{fig:rho}
\end{figure}

The selection of $\theX\to\pip\pim\jpsi$ candidates is based on the reconstruction of  $\decay{\Bu}{\Kp\pip\pim\jpsi(\to\mu^+\mu^-)}$ decays, which provides for efficient background suppression and optimal dipion mass resolution. Both muon candidates are identified by the muon detector. 
The dimuon mass must be consistent with the known $\jpsi$ mass~\cite{PDG2020}, and all five final state particles must form a good-quality vertex significantly displaced from the closest primary $pp$ interaction vertex (PV).
The hadron candidate most likely to be a kaon is selected as the kaon candidate. 
Each hadron must have a significant impact parameter with respect to any PV.
To remove multiple entries per event, the $\Bu$ candidate with the largest scalar sum of the hadron and $\jpsi$ candidate transverse-momenta is selected.
To improve mass resolution, the $\Bu$ candidates are kinematically constrained to point to the closest PV and reproduce the known $\jpsi$ mass~\cite{Hulsbergen:2005pu}.
The mass of these candidates must be consistent with the known $\Bu$ mass \cite{PDG2020}, which is then also included among the kinematic constraints.
The resulting $\mppj$ distribution 
(Fig.~\ref{fig:xmass_fit}) is fit with a signal (background) shape modelled by 
a double-sided Crystal Ball~\cite{Skwarnicki:1986xj} (quadratic) function, yielding
$6788\pm117$ 
$\theX\to\pip\pim\jpsi$ decays with 
a $\theX$ mass resolution of $\sigma_m=2.66\pm0.09$\mev (natural units are used throughout).
The signal purity is $77\%$ within $\pm2\,\sigma_m$ around the  
$\theX$ mass.
The dominant source of background is from
$\Bu$ decays to $\jpsi$ meson and excited kaons ($K^{*+}$), which decay to $\Kp\pip\pim$.
The dipion mass distribution is obtained by two-dimensional unbinned fits of the $\theX$ signal yields to the $(\mppj,\mpp)$ data in $\mpp$ intervals. The signal shape in $\mppj$ is fixed to the global fit result of Fig.~\ref{fig:xmass_fit}, while the background shape may vary in each $\mpp$ interval. The signal and the background $\mppj$ shapes are then multiplied by a two-body phase-space factor, \ie the $\jpsi$ momentum in the $\theX$ rest frame ($\pjpsi$), which depends on both $\mppj$ and $\mpp$. The resultant shapes are normalized to unity using their integral over the fitted phase-space range. The obtained data points are shown in Fig.~\ref{fig:rho}.

Signal simulation is used to obtain the $\mpp$ dependence of the dipion mass resolution and of the reconstruction efficiency.  
In the simulation, $pp$ collisions are generated using \pythia~\cite{Sjostrand:2007gs,*Sjostrand:2006za} with a specific \lhcb configuration~\cite{LHCb-PROC-2010-056}.
Decays of unstable particles
are described by \evtgen~\cite{Lange:2001uf}, in which final-state
radiation is generated using \photos~\cite{davidson2015photos}.
In particular, $\Bu\to\Kp\theX$, $\theX\to\jpsi\rhoz$, $\rhoz\to\pi^+\pi^-$, $\jpsi\to\mu^+\mu^-$ signal decays are simulated using the helicity model where the $\theX$ decays via an $S$-wave, which describes the angular distributions in the data well~\cite{LHCb-PAPER-2015-015}. 
 When integrating over $\mpp$,
 the generated $\mpp$ distribution is weighted to reproduce the observed distribution.
The interaction of the generated particles with the detector, and its response,
are implemented using the \geant
toolkit~\cite{Allison:2006ve, *Agostinelli:2002hh} as described in
  Ref.~\cite{LHCb-PROC-2011-006}. 
The underlying $pp$ interaction is reused multiple times, with an independently generated signal decay for each event~\cite{LHCb-DP-2018-004}.
The transverse-momentum distribution of the signal $\Bu$ is weighted to match the data.
The relative signal reconstruction efficiency obtained from the simulation is well described by a quadratic function, 
\hbox{
$\epsilon(\mpp)\!=\!0.966\!+\!1.345\!\times\!10^{-3}\,(\mpp\!-\!700)\!+\!1.607\!\times\!10^{-6}\,(\mpp\!-\!700)^2$}, and 
the dipion mass resolution by \hbox{$\sigma(\mpp)\!=\! 2.39\,(1\!-\!\exp(-\mpp\!/220.3))\!-\!5.4\,\exp(-\mpp\!/220.3)$}, where $\mpp$ and $\sigma(\mpp)$ are in~\mev. The simulation underestimates the $\theX$ and $\Bu$ mass resolution by 6\% and 14\%, respectively. Therefore, $\sigma(\mpp)$ is scaled up by $f_m=1.06$, which is varied from $1.00$ to $1.14$ to assess a systematic uncertainty.
All theoretical probability density functions fit to the data, $\PDF(\mpp)$, are multiplied by the relative efficiency function, $\epsilon(\mpp)$ and convolved with a Gaussian distribution characterized by a root-mean-square value of $f_m \sigma(\mpp)$.

The matrix element, $\M$, describing the three-body decay, \hbox{$\theX\to\pi^+\pi^-\jpsi$}, is related to 
$\PDF(\mpp)$ via a scaling ($S$) and the phase-space factors, 
\hbox{$\PDF(\mpp)=S\,\pjpsi\,\ppi\,|\M|^2$},
where $\ppi$ is the pion momentum in the rest frame of the $\pip\pim$ system. The scaling factor $S$ is a nuisance parameter.

Following the previous analyses~\cite{CDF:2005cfq,Belle:2011vlx}, 
the $\rhoz$ component is first parameterized as  
a Breit--Wigner (BW) amplitude,
\begin{equation}
    \M=\mathrm{BW}(s\,|\,m_\rho,\Gamma_\rho)= 
    \frac{m_\rho\Gamma_\rho
    \sqrt{{B_1(\ppi)}/{B_1(\ppi^\rho)}}
    }{{m_\rho}^2-s-i\,m_\rho\Gamma(s)},\quad 
    \Gamma(s)=\Gamma_\rho\,\frac{\ppi}{\ppi^\rho}\,\frac{m_\rho}{\sqrt{s}}\,
    \frac{B_1(\ppi)}{B_1(\ppi^\rho)},
    \label{eq:bw}
\end{equation}
where $s\equiv\mpp^2$, $\ppi^\rho = \ppi(m_\rho)$,
$m_\rho=775.26$\mev, $\Gamma_\rho=147.4\mev$~\cite{PDG2020}  
and \hbox{$B_1(p) = p^2 / [(1+(R\,p)^2)]$} is the Blatt--Weisskopf barrier factor for the $P$-wave decay of a vector particle to $\pi^+\pi^-$, and contains an effective hadron-size parameter $R$.
With $R$ adjusted to 1.45$\gev^{-1}$, a complex phase of the BW amplitude varies in the fitted $\mpp$ range within one degree of the isovector $\pip\pim$ $P$-wave parametrization extracted from the scattering data by the phenomenological analysis of Ref.~\cite{Garcia-Martin:2011iqs}.
A similar value of 1.5$\gev^{-1}$ was used in the previous analyses~\cite{CDF:2005cfq,Belle:2011vlx}. 
The $\chi^2$ fit, with $S$ as the only fit parameter, fails to describe the data as shown in Fig.~\ref{fig:rho}, with a $\chi^2$ value per number of degrees of freedom ($\chindf$) equal to $366.6/34$. 
A large disagreement between the data and the $\theX\to\rhoz\jpsi$ amplitude at high dipion masses was missed in the previous comparisons with the
$\theX\to\rhoz\jpsi$ simulation
(see Fig.~S2 in the supplementary material of Ref.~\cite{LHCb-PAPER-2015-015}; see also 
Refs.~\cite{CMS:2013fpt,ATLAS:2016kwu}), because 
the \evtgen~\cite{Lange:2001uf} model does not correctly simulate the impact of
the phase-space factors (here $\pjpsi\ppi$) on resonant shapes. As a consequence, the large $\rhoz-\omega$ interference in the data (see below) was mistakenly interpreted as a part of the $\rhoz$ resonance itself.

Adding an $\omega$ contribution via the BW sum model with $\M\!=\! \mathrm{BW}(s|m_\rho,\Gamma_\rho)\!+\!
    a\, \exp(i\,\phi)\,\mathrm{BW}(s|m_\omega,\Gamma_\omega)$, where $a$ and $\phi$ are the amplitude and phase of the $\omega$ component relative to the $\rhoz$ term~\cite{CDF:2005cfq,Belle:2011vlx}, improves the fit quality substantially, $\chindf=102.9/33$, 
    yet not enough to be acceptable.
Summing BW amplitudes results in matrix elements that are not unitary, violating first principles of scattering theory. 
A two-channel $K$-matrix ($K$) model \cite{Aitchison:1972ay}, coupling the $\pi^+\pi^-$ and $\pi^+\pi^-\pi^0$ channels via an $\omega$ contribution, resolves this issue,
\begin{equation}
    K =
    \frac{1}{{m_\rho}^2-s} \left( 
    \begin{array}{cc}
    \grppsq & 0 \\
    0         & 0 \\
    \end{array}
    \right) 
    +
    \frac{1}{{m_\omega}^2-s} \left( 
    \begin{array}{cc}
    \goppsq   & \gopp\,\goppp \\
    \gopp\,\goppp & \gopppsq \\
    \end{array}
    \right),
\label{eq:kmatrix}
\end{equation}
where $g$ are the coupling constants described later.
The $g_{\rho\to3\pi}^2$ coupling is 4--5 orders of magnitude smaller than $g_{\rho\to2\pi}^2$~\cite{PDG2020,Achasov:2003ir,Benayoun:2009im}
and has been neglected here.
The $T$-matrix is obtained from
$\hat{T} = \left[ 1 - i\,K\,\varrho\right]^{-1}\,K$,
where the phase-space matrix $\varrho$ is diagonal, $\varrho=\mathrm{diag}(\varrho_{2\pi}(s),\varrho_{3\pi}(s))$,
and is detailed in  the supplemental material.
The decay amplitude is given by 
\begin{equation}
 \M = \hat{A}_{2\pi} \sqrt{B_1(\ppi)},
\label{eq:mcoupled}
\end{equation}
where $\hat{A}_{2\pi}\equiv\alpha_{2\pi}\hat{T}_{2\pi,2\pi}+\alpha_{3\pi}\hat{T}_{2\pi,3\pi}$, 
and the elements of the production $Q$-vector ($\alpha_{2\pi}$, $\alpha_{3\pi}$) are real~\cite{Chung:1995dx}.
The coupling constants are fully determined from other experiments~\cite{PDG2020},
\begin{eqnarray}
\grppsq&=&m_\rho\,\Gamma_\rho/\varrho_{2\pi}(m_\rho^2),\\ \gopppsq &=&m_\omega\,\Gamma_\omega \BR(\omega\to\pip\pim\piz)/\varrho_{3\pi}(m_\omega^2),\\ 
\goppsq&=&m_\omega\,\Gamma_\omega\BR(\omega\to\pi^+\pi^-)/\varrho_{2\pi}(m_\omega^2).
\end{eqnarray}
Numerically, $\goppsq/\grppsq\approx0.0009$, while $\gopp\,\goppp/\grppsq\approx0.01$. Thus, the diagonal $\omega$ coupling to two pions can be neglected in comparison with the off-diagonal coupling. 
In this approximation, equivalent to the full formalism given the precision of this analysis,
\begin{equation}
 \hat{T}_{2\pi,2\pi}\approx\frac{\grpp^2}{{m_\rho}^2-s-i\,\grpp^2\varrho_{2\pi}(s)}, 
 \label{eq:T11}
\end{equation}
which is the $\rhoz$ Breit--Wigner amplitude (Eq.~\ref{eq:bw}) and $\alpha_{2\pi}$ is the $\rhoz$ production factor in the $\theX$ decay.
A mild dependence of $\alpha_{2\pi}$ on $s$ is possible, since $\alpha_{2\pi}(s)$ must be analytic and cannot change much within the resonance widths (see \eg~Ref.~\cite{JPAC:2017dbi}). 
Using Chebyshev polynomials ($C_n$) to express the dependence: 
$\alpha_{2\pi}(s) = \sum_{n=0}^{n=N} P_n\, C_n(\hat{s})$, 
where 
$\hat{s} \equiv 2\,(s-s_{\rm min})/(s_{\rm max}-s_{\rm min})-1$, 
$s_{\rm min}=(380\mev)^2$ and $s_{\rm max}=(775\mev)^2$.
The polynomial coefficients $P_n$ are determined from a fit to the data, except 
for $P_0$, which is set to unity as the normalization choice. For  $\alpha_{2\pi}(s)$ to have the expected theoretical behavior, the series must converge, $|P_{n+1}|<|P_n|$ and $N$ should be kept small. 

The $\omega$ contribution enters via the element,
\begin{equation}
 \hat{T}_{2\pi,3\pi}\approx\frac{\gopp\,\goppp\,(m_\rho^2-s)}{({m_\rho}^2-s-i\,\grpp^2\varrho_{2\pi}(s))({m_\omega}^2-s-i\,\gopppsq\varrho_{3\pi}(s))}. 
 \label{eq:T12}
\end{equation}
This term approaches zero at the bare $\rhoz$ pole mass, which complicates probing the $\omega$ contribution,
$m_\omega=782.66\pm0.13$\mev and $\Gamma_\omega=8.68\pm0.13\mev$~\cite{PDG2020}.
This zero is an artifact of the $K$-matrix approach, rather than an expectation from scattering theory.
To remove it and restore a more physical behavior of the $\omega$ term, 
\begin{equation}
    \alpha_{3\pi} = a_\omega\, \frac{{m_\omega}^2-{m_\rho}^2}{{m_\rho}^2-s}
    \label{eq:alpha2},
\end{equation}
is set.
The constant term ${m_\omega}^2-{m_\rho}^2$ is introduced above to make $a_\omega$ express the ratio of $\omega/\rho$ amplitudes at the $\omega$ pole.
The value of $a_\omega$ is determined by the fit to the data.  
Using Eqs.~\ref{eq:T11}, \ref{eq:T12} and \ref{eq:alpha2}, 
with $\alpha_{2\pi}$ constant ($N=0$), is the common method to describe $\rhoz-\omega$ interference in analyses of  the $\pi^+\pi^-$ system~\cite{Goldhaber:1969dp}. The exact $K$-matrix formulae are used here.
In view of Eq.~\ref{eq:T11}, models with $\alpha_{3\pi}=0$ are interpreted as containing a $\rhoz$ component only. 
The following integrals are calculated to quantify a relative rate of the $\omega$ contribution:
$I_\mathrm{tot} = \int \PDF(\mpp) d\mpp$,
$I_{\rho} =\int  \PDF(\mpp\,|\,\alpha_{3\pi}=0) d\mpp$ and 
$I_{\omega} = \int  \PDF(\mpp\,|\,\alpha_{2\pi}=0) d\mpp$,
where $\PDF(\mpp)$ is neither convolved with the mass resolution, nor multiplied by the efficiency function. The integration is performed over the full phase space.
To quantify the overall impact of $\omega$ on the total rate, including $\rhoz-\omega$ interference effects, the ratio $\Rall\equiv 1-{I_\rho}/{I_\mathrm{tot}}$ is defined.
The traditional $\omega$ fit fraction is given by $\Rzero\equiv {I_\omega}/{I_\mathrm{tot}}$. 
Finally, to probe the ratio of the $\theX$ isospin conserving to isospin violating couplings, the ratio $\Riso\equiv {I_\omega}/{I_\rho}$ is introduced.
Using the $\Bu\to\Kp\theX$ simulation,
the extraction of the dipion mass distribution and subsequent fit with the $K$-matrix model, with the efficiency and the mass resolution corrections included, 
are verified to a precision an order of magnitude better than the total systematic uncertainties reported below. 

\begin{figure}[tb]
  \begin{center}
    \includegraphics[width=0.7\linewidth]{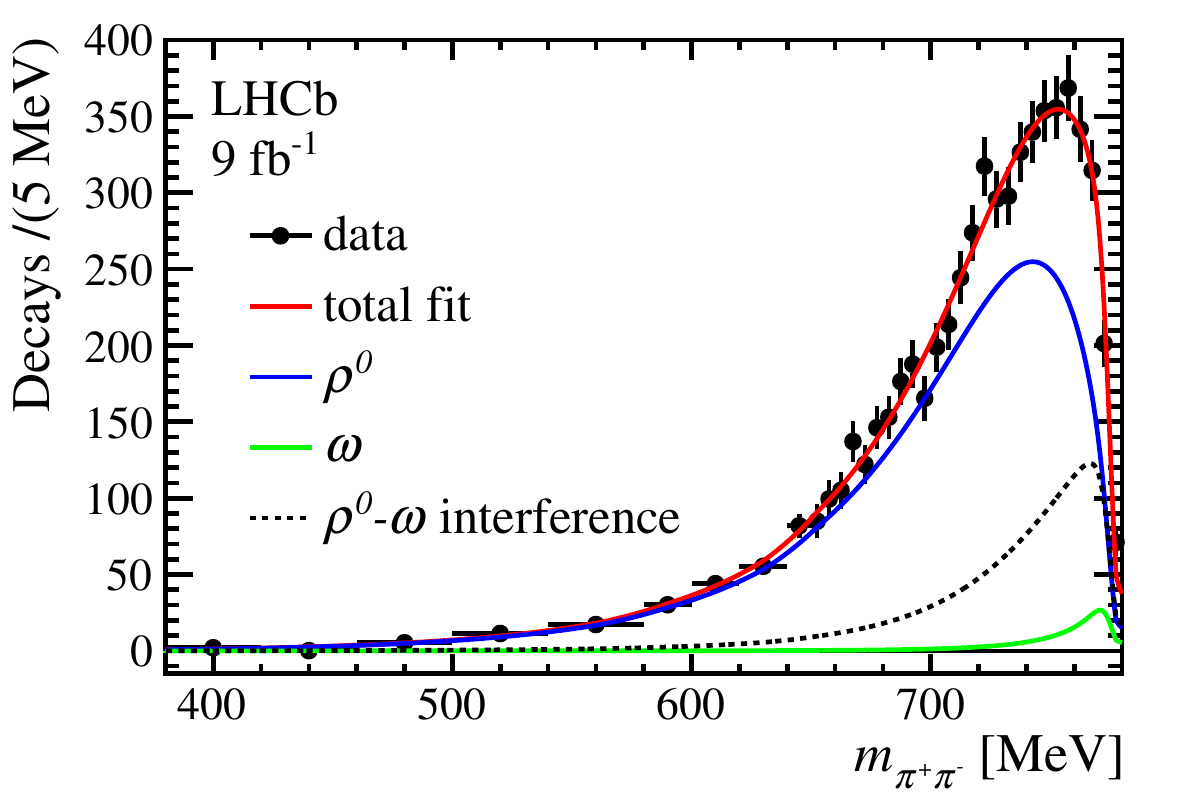}
    \vspace*{-0.5cm}
  \end{center}
  \caption{
    \small 
        Distribution of $\mpp$ in $\theX\to\pip\pim\jpsi$ decays,
        fit with the default $K$-matrix model.}
  \label{fig:default}
\end{figure}

Allowing the $\omega$ term in the $K$-matrix model with the constant $\alpha_{2\pi}$ improves the fit quality to  $\chindf=55.1/33$, 
resulting in a significance for the $\omega$ contribution of $n_\sigma=(\chi^2_{a_\omega=0}-\chi^2_{a_\omega\ne0})^{1/2}=17.7\sigma$~\cite{Wilks:1938dza}. However, the $p$-value of the fit is marginal (0.9\%).
Introducing a term linear in $s$ to $\alpha_{2\pi}(s)$ improves the fit with $\chindf=24.7/32$, $p$-value $=82\%$, and
a small slope coefficient, $P_1=0.23\pm0.05$. This fit, shown in Fig.~\ref{fig:default}, is taken as the default model 
and yields $n_\sigma=8.1\sigma$, $\Rall=0.214\pm0.023$, $\Rzero=0.0193\pm0.0044$,  $\Riso=0.0246\pm0.0062$ and $a_\omega=0.208\pm0.024$.
The $\omega$ contribution is at the level of $\Rzero\approx2\%$ as expected from the $\theX\to\omega\jpsi$, $\omega\to\pip\pim\piz$ measurements (see above). The impact on the overall $\theX\to\pip\pim\jpsi$ rate given by $\Rall$ is an order of magnitude larger, enhanced by the $\rhoz-\omega$ interference. 
Allowing a quadratic term in $\alpha_{2\pi}(s)$ lowers the $p$-value of the fit to $78\%$ with $\chindf=24.6/31$, indicating that too much freedom is added to the model.
The $P_2$ coefficient is small and consistent with zero, $P_2=0.016\pm0.047$, while all other fit results remain consistent with the default fit, $P_1=0.21\pm0.07$, $\Rall=0.206\pm0.035$, $\Rzero=0.0178\pm0.0062$, $\Riso=0.0225\pm0.088$ and $a_\omega=0.197\pm0.042$.
The significance of the $\omega$ contribution remains very high, $n_\sigma=5.5\sigma$, and is now underestimated, since the null hypothesis ($a_\omega=0$) gives an unphysical polynomial correction (the quadratic term, $P_2=0.17\pm0.02$, comparable to, and more significant than the linear term, $P_1=0.16\pm0.05$).

A number of analysis variations have been performed to evaluate systematic uncertainties on the $\omega$ fraction, as summarized in Table~\ref{tab:sys}.
In the dipion mass extraction from the data, Gaussian functions are used for the $\theX$ shape in $\mppj$ data, rather than Crystal Ball functions.
A cubic polynomial replaces the quadratic function in the relative efficiency parametrization.

Data-driven corrections to the simulation of the hadron identification are implemented, as a cross-check. 
To check further the relative efficiency simulation and the background subtraction, a tighter data selection is implemented with a Boosted Decision Tree (BDT) classifier~\cite{Hocker:2007ht}, using inputs from hadron identification variables, the $\Bu$ vertex quality, the PV impact parameters of the final-state particles and the $\Bu$ candidate, the hadron transverse-momenta and the $\Bu$ flight distance.  The tighter selection reduces the combinatoric background under the $\Bu$ peak in the $m_{\Kp\pip\pim\jpsi}$ distribution from 9.4\%\ to 3.1\%, resulting in an overall background reduction under the $\theX$ peak in the $\mppj$ distribution from 23\%\ to 17\%. At the same time, the $\theX$ signal yield is reduced by only $0.4\%$.  

\begin{table}[hbtp]
    \centering
    \renewcommand{\arraystretch}{1.5}
    \caption{Results for $\omega$ fractions with systematic uncertainties. See text for a description of the individual entries.}
    \label{tab:sys}    
    \resizebox{\columnwidth}{!}{
    \begin{tabular}{l|c|c|c|c|c|c}
      Fit type   &  $\chindf$ & $p$-value & $\Rall$ & $\Rzero$ & $\Riso$ & $n_\sigma$  \\
                 \hline\hline
      Default    & $24.7/32$  & $0.82$  & $0.214\pm0.023$ & $0.019\pm0.004$ & $0.025\pm0.006$ 
                                 & $8.1\sigma$  \\
                \hline                 
      $P_2\ne0$    & $24.6/31$  & $0.78$  & $0.206\pm0.035$ & $0.018\pm0.006$ & $0.023\pm0.009$ 
                                 & $5.5\sigma$  \\                                 
    Gaussian $\theX$ & $20.0/32$ & $0.95$ & $0.194\pm0.024$ & $0.016\pm0.004$ & $0.020\pm0.006$ & $7.3\sigma$  \\ 
    cubic $\epsilon(m_{\pip\pim})$ & $24.5/32$ & $0.83$  & $0.221\pm0.023$ & $0.021\pm0.005$ & $0.027\pm0.007$ 
                                 & $8.1\sigma$              \\             
    had.ID corrections & $24.6/32$ & $0.82$  & $0.214\pm0.023$ & $0.019\pm0.004$ &
    $0.025\pm0.006$                             & $8.1\sigma$              \\
    BDT selection & $24.6/32$ & $0.82$  & $0.207\pm0.022$ & $0.018\pm0.004$ & $0.023\pm0.006$ 
                                 & $7.9\sigma$              \\
  $\sigma(\mpp)\times1.0$    &$26.6/32$  & $0.74$ & 
   $0.213\pm0.023$ & $0.019\pm0.004$ & $0.025\pm0.006$
                                 & $8.1\sigma$              \\
    $\sigma(\mpp)\times1.14$    &$22.6/32$  & $0.89$ & $0.215\pm0.023$ & $0.020\pm0.004$ & $0.026\pm0.006$ 
                                 & $8.1\sigma$   \\
    $\mpp<775$ \mev &$18.0/31$  & $0.97$ & $0.196\pm0.024$ & $0.016\pm0.004$ &
          $0.021\pm0.006$                       & $7.1\sigma$   \\
                  $\cos\theta_X<0$ & $26.9/32$ & $0.72$ & $0.211\pm0.035$  & $0.019\pm0.007$ &
   $0.024\pm0.010$
                                 & $5.2\sigma$               \\
   $\cos\theta_X>0$ & $42.2/32$ & $0.11$ & $0.217\pm0.030$  & $0.021\pm0.006$ & $0.027\pm0.009$
                                 & $4.2\sigma$              \\
   NR prod.\ of $2\pi$ & $24.7/32$ & $0.82$ & $0.214\pm0.022$  & $0.019\pm0.004$ & $0.025\pm0.006$
                                 & $8.1\sigma$             \\
   $D$-wave free & $24.5/31$ &  $0.79$ &
   $0.210\pm0.029$ & $0.017\pm0.005$ & $0.021\pm0.007$ &
                         $7.8\sigma$ \\
  $D$-wave fixed at $4\%$ & $24.5/32$ &  $0.82$ &
   $0.208\pm0.023$ & $0.018\pm0.004$ & $0.023\pm0.006$ &
                         $7.9\sigma$ \\
   $\rho'$  & $25.1/32$ & $0.80$  & $0.209\pm0.023$ & $0.018\pm0.004$ &
        $0.024\pm0.006$                         & $8.1\sigma$              \\
             $R_\mathrm{prod}=0$ $\gev^{-1}$ &$24.7/32$  &$0.82$ & $0.209\pm0.023$ & $0.019\pm0.004$ & $0.024\pm0.006$
                                 & $7.9\sigma$ \\
   $R_\mathrm{prod}=30$ $\gev^{-1}$ &$24.6/32$  &$0.82$ & $0.229\pm0.022$ & $0.021\pm0.004$ & $0.028\pm0.006$
                                 & $8.7\sigma$   \\                             
    $R=1.3$ $\gev^{-1}$ &$24.7/32$  & $0.82$  & $0.216\pm0.022$  & $0.020\pm0.004$ & $0.026\pm0.006$
                                 & $8.2\sigma$               \\
   $R=1.6$ $\gev^{-1}$ &$24.7/32$  &$0.82$ & $0.212\pm0.023$ & $0.019\pm0.004$ &
         $0.025\pm0.006$                        & $8.0\sigma$              \\

    GS model  & $24.8/32$ &  $0.81$  &  
    $0.221\pm0.024$  &  $0.021\pm0.005$ &
        $0.028\pm0.007$         & $7.8\sigma$  \\
                             
                                 \hline\hline
   Summary       &         &          &  {\small $0.214\!\pm\!0.023\!\pm\!0.020$}  &   
   {\small $0.019\!\pm\!0.004\!\pm\!0.003$}  &
   {\small $0.025\!\pm\!0.006\!\pm\!0.005$}
   & 
   $>7.1\sigma$     \\

    \end{tabular}
    }
\end{table}

To evaluate uncertainties due to the mass resolution used in the fit, the scaling factor $f_m$ is varied and the fit range is reduced to exclude the last interval of the $\mpp$ distribution (775--780\mev), 
since this falls beyond the phase-space limit, and therefore its content is very sensitive to $\sigma(\mpp)$.

To check for interference effects between $\Bu\to\Kp\theX$ and $\Bu\to K^{*+}\jpsi$ decays, the data are split into two subsamples depending on the sign of $\cos\theta_X$, where $\theta_X$ is the $\theX$ helicity angle, the angle between the $\Kp$ and $\jpsi$ momenta in the $\theX$ rest frame. The composition of $K^{*+}$ resonances is different in each subsample.

As a variation of the production model,
non-resonant (NR) terms are added to the production vector, 
$(\hat{A}_{2\pi},\hat{A}_{3\pi})=$ $[1-i\,K\,\varrho]^{-1}$ $\def\1#1#2{(#1,#2)}%
[K\1{\alpha_{2\pi}}{\alpha_{3\pi}} + \1{f_{2\pi}}{f_{3\pi}}]$.
Without $\theX\to3\pi\jpsi$ data in the fit, the NR production parameter $f_{3\pi}$ cannot be probed, thus it is set to zero.
A good-quality fit is obtained with constant $\alpha_{2\pi}$, which gives $f_{2\pi}=(-9.7\pm1.6)\times10^{-7}$.

The default fit assumes an $S$-wave $\theX$ decay. 
When a $D$-wave component is allowed in the fit, multiplying $|\M|^2$ by
$1 + {A_D}^2 \, {B_2(\pjpsi)}/{B_2(\pjpsi^\rho)}$, where $\pjpsi^\rho\equiv\pjpsi(m_\rho)$ and 
$B_2(p)  \equiv  p^4/{(9 + 3\, (R\,p)^2 + (R\,p)^4)}$, 
the $A_D$ parameter is consistent with zero, \mbox{$0.13\pm0.41$}. Tuning $A_D$ to $0.176$ produces a 4\%\ $D$-wave fraction, equal to the upper limit from studies of the angular correlations~\cite{LHCb-PAPER-2015-015}.

Since the hadron size parameter $R$ is tuned to the scattering data, which automatically includes any $\rhoz$ excitations, there is no strong motivation to include the $\rho(1450)$ pole~\cite{PDG2020} ($\rho'$) in the $K$-matrix model. When included, the fit quality is slightly reduced.

 The size parameter $R$ used in $B_1(\ppi)$ of Eq.~\ref{eq:mcoupled} could be related to the $\theX$ size, rather than the $\rhoz$ or $\omega$ sizes. Thus, fit variations are tried in which it has an independent value, $R_\mathrm{prod}$, varied from zero to 30$\gev^{-1}$.  
The $R$ value itself is varied within the range 1.3--1.6$\gev^{-1}$.
An alternative model of the $\rhoz$ shape, that does not include an $R$-dependent Blatt--Weisskopf form factor, is provided by the Gounaris--Sakurai (GS) formula~\cite{Gounaris:1968mw},
and was utilized by the BaBar collaboration
to describe high statistics $e^+e^-\to\pi^+\pi^-(\gamma)$ data~\cite{Lees:2012cj}. 
As discussed in the supplemental material,
when applying this prescription with $\rhoz$, $\rho'$ and $\omega$ contributions, an excellent fit to the data is obtained, $\chindf=24.8/32$, $p$-value $=81\%$,
matching the fit quality of the default fit.

Total systematic uncertainties, given at the bottom of Table~\ref{tab:sys}, are set to cover the maximal deviation from the default fit results, and are comparable to the statistical uncertainties. 
The lowest $\omega$ significance is $7.1\sigma$, excluding the subsamples and $P_2\ne0$ fit as discussed above.
This is a more significant observation of $\theX\to\omega\jpsi$ decays than achieved using the dominant 
$\omega\to\pip\pim\piz$ decay channel.

In addition to the resonant coupling constants, the limited phase space in $\theX\to\pip\pim\jpsi$ decays has a large impact on the $\Riso$ value capturing a much smaller fraction of the $\omega$ resonance than for the $\rhoz$ resonance.  
To probe the resonant coupling constants, the phase space can be artificially extended to contain both resonance peaks by 
setting $m_{\theX}=4000$ \mev, as illustrated in the supplemental material. Integrating the default $\PDF(\mpp)$ model in the extended phase space, ${\Riso}'=0.18\pm0.05$ is obtained and then used to deduce the ratio of the isospin violating to isospin conserving $\theX$ couplings,
\begin{equation}
    \frac{g_{\theX\to\rhoz\jpsi}}{g_{\theX\to\omega\jpsi}} =
    \sqrt{ \frac{\BR(\omega\to\pip\pim)}{{\Riso}'} } =
    0.29\pm0.04.
    \notag
\end{equation}
This value is an order of magnitude larger than expected for pure $\ccbar$ states, as exemplified by the corresponding value for the $\psi(2S)$ state~\cite{PDG2020},
\begin{equation}
\frac{g_{\psi(2S)\to\piz\jpsi}}{g_{\psi(2S)\to\eta\jpsi}}
=
\sqrt{
\frac{\BR(\psi(2S)\to\piz\jpsi)}{\BR(\psi(2S)\to\eta\jpsi)}
\,
\frac{{p_\eta}^3}{{p_\piz}^3}}
=0.045\pm0.001,
\notag
\end{equation} 
where $p_\eta$ and $p_\piz$ are the breakup momenta~\cite{Hanhart:2011tn}.
Therefore, the $\theX$ state cannot be a pure charmonium state.
Large isospin violation is naturally expected 
in models in which the $\theX$ state has a significant $D\Dstarb$ component, 
either as constituents (\ie in the \lq\lq molecular model") or generated dynamically in the decay~\cite{Tornqvist:2003na,Tornqvist:2004qy,Voloshin:2003nt,Swanson:2003tb,Suzuki:2005ha,Coito:2010if,Li:2012cs,Takeuchi:2014rsa,Wu:2021udi,Meng:2021kmi}.
The proximity of the $\theX$ mass to the $D^0\Dstarzb$ threshold, enhances such contributions over $D^+D^{*-}$ combinations, which are 8\mev heavier.
It has also been suggested in compact tetraquark models that two neutral states could be degenerate and mix, giving rise to large isospin violation in $\theX$ decays~\cite{Terasaki:2007uv,Maiani:2017kyi,Maiani:2020zhr}.

In summary, the $\rhoz$ and $\omega$ contributions to  $\theX\to\pip\pim\jpsi$ decays are resolved for the first time  using a much larger data sample than previously available. 
Through $\rhoz-\omega$ interference, the $\omega$ contribution accounts for $(21.4\pm2.3\pm2.0)\%$ of the total rate, or equivalently, $\BR(\theX\to\rhoz\jpsi)/\BR(\theX\to\pip\pim\jpsi)=(78.6\pm2.3\pm2.0)\%$.
Excluding interference effects, the $\omega$ contribution, $(1.9\pm0.4\pm0.3)\%$, is found to be consistent with, but more precise than the previous $\theX\to\omega\jpsi$ measurements utilizing $\omega\to\pip\pim\piz$ decays~\cite{Abe:2005ax,delAmoSanchez:2010jr,BESIII:2019qvy}. 
The isospin violating $\rhoz$ contribution, quantified for the first time with proper subtraction of the $\omega$ contribution, is an order of magnitude too large  for $\theX$ to be a pure charmonium state.

This analysis also serves as an excellent illustration for why a simple sum of $\rhoz$ and $\omega$ Breit--Wigner amplitudes should not be used to describe interfering resonances. Such a model fails to describe the data, while the $K$-matrix approach, which respects unitarity, provides an excellent description of the data and gives numerical results that are consistent with the previous $\theX\to\omega\jpsi$ measurements utilizing $\omega\to\pip\pim\piz$ decays.   

\section*{Acknowledgements}
%
%
\noindent We express our gratitude to our colleagues in the CERN
accelerator departments for the excellent performance of the LHC. We
thank the technical and administrative staff at the LHCb
institutes.
We acknowledge support from CERN and from the national agencies:
CAPES, CNPq, FAPERJ and FINEP (Brazil); 
MOST and NSFC (China); 
CNRS/IN2P3 (France); 
BMBF, DFG and MPG (Germany); 
INFN (Italy); 
NWO (Netherlands); 
MNiSW and NCN (Poland); 
MEN/IFA (Romania); 
MICINN (Spain); 
SNSF and SER (Switzerland); 
NASU (Ukraine); 
STFC (United Kingdom); 
DOE NP and NSF (USA).
We acknowledge the computing resources that are provided by CERN, IN2P3
(France), KIT and DESY (Germany), INFN (Italy), SURF (Netherlands),
PIC (Spain), GridPP (United Kingdom), 
CSCS (Switzerland), IFIN-HH (Romania), CBPF (Brazil),
Polish WLCG  (Poland) and NERSC (USA).
We are indebted to the communities behind the multiple open-source
software packages on which we depend.
Individual groups or members have received support from
ARC and ARDC (Australia);
Minciencias (Colombia);
AvH Foundation (Germany);
EPLANET, Marie Sk\l{}odowska-Curie Actions and ERC (European Union);
A*MIDEX, ANR, IPhU and Labex P2IO, and R\'{e}gion Auvergne-Rh\^{o}ne-Alpes (France);
Key Research Program of Frontier Sciences of CAS, CAS PIFI, CAS CCEPP, 
Fundamental Research Funds for the Central Universities, 
and Sci. \& Tech. Program of Guangzhou (China);
MEFP, GVA, XuntaGal, GENCAT and Prog.~Atracci\'on Talento, CM (Spain);
SRC (Sweden);
the Leverhulme Trust, the Royal Society
 and UKRI (United Kingdom).


\clearpage


\clearpage
\newpage

{\normalfont\bfseries\boldmath\huge
\begin{center}
 \papertitle  \\
\vspace{0.05in}
{ \it \Large Supplemental material}\\
\vspace{0.05in}
\end{center}
}

\setcounter{equation}{0}
\setcounter{figure}{0}
\setcounter{table}{0}
\setcounter{section}{0}
\renewcommand{\theequation}{S\arabic{equation}}
\renewcommand{\thefigure}{S\arabic{figure}}
\renewcommand{\thetable}{S\arabic{table}}
\newcommand\ptwiddle[1]{\mathord{\mathop{#1}\limits^{\scriptscriptstyle(\sim)}}}

\section{$K$-matrix phase-space matrix elements}
A notation in which the Blatt--Weisskopf barrier factors ($B_l$) are integrated with the phase-space matrix is used in this Letter,  
\begin{equation}
\varrho_{2\pi}(s)  =  \frac{2\ppi}{\sqrt{s}}\,B_1(\ppi).
\end{equation}
Since the $\omega$ meson decays to three pions via a $P$-wave $\rho\pi$ decay,
\begin{eqnarray}
\varrho_{3\pi}(s) & = &\int_{(2m_\pi)^2}^{(\sqrt{s}-m_\pi)^2}
d\sigma 
\frac{2\ppi(\sigma)}{\sqrt{\sigma}}
\frac{B_1(\ppi(\sigma))}{({m_\rho}^2-\sigma)^2+(m_\rho\,\Gamma_\rho)^2}
\frac{2p_3'(s,\sigma)}{\sqrt{s}}\,
 B_1(p_3'(s,\sigma))
 \label{eq:rhotwo}\\
 p_3'(s,\sigma) & = &
\frac{\sqrt{(s-(\sqrt{\sigma}+m_\pi)^2)(s-(\sqrt{\sigma}-m_\pi)^2)}}{2\sqrt{s}}.
\end{eqnarray}

\section{Gounaris--Sakurai resonant $\mpp$ shape}

An alternative model to the $R$-dependent Blatt--Weisskopf form factor
in the Breit--Wigner amplitude is provided by the Gounaris--Sakurai formula~\cite{Gounaris:1968mw},
\begin{align} \label{eq:GS.rho}
\mathrm{BW}_\rho^\text{GS}(s\,|\,m_\rho,\Gamma_\rho) =& 
\frac{m_\rho^2\left[ 1+ d(m_\rho) \Gamma_\rho/m_\rho \right]}
{m_\rho^2-s+f(s,m_\rho,\Gamma_\rho)-i\, m_\rho\Gamma(s,m_\rho,\Gamma_\rho)},
\end{align}
where,
\begin{align}
\Gamma(s,m,\Gamma_0) = & \Gamma_0 \frac{m}{\sqrt{s}}\left[ \frac{\ppi(s)}{\ppi(m^2)}\right]^3,\\
d(m) = & \frac{3}{\pi}\frac{m_\pi^2}{\ppi^2(m^2)}
\log \left[ \frac{m + 2\ppi(m^2)}{2m_\pi}\right]
+\frac{m}{2\pi \ppi(m^2)}
-\frac{m_\pi^2 m }{\pi \ppi^3(m^2)},\\
f(s,m,\Gamma_0) = & \frac{\Gamma_0 m^2}{\ppi^3(m^2)}
\left[
\ppi^2(s) \left[ h(s)-h(m^2) \right] 
+(m^2-s)\ppi^2(m^2)h'(m^2)
 \right],\\
 h(s) = & \frac{2}{\pi}\frac{\ppi(s)}{\sqrt{s}}
 \log \left[\frac{\sqrt{s}+2\ppi(s)}{2m_\pi}\right],
\end{align}
and $h'(s)$ is the derivative of $h(s)$, with respect to $s$, calculated numerically.
To include the $P$-wave momentum barrier in the $\rhoz$ decay, the matrix element is set to
\hbox{$\M={\ppi(s)}/{\ppi(m_\rho)}\,\mathrm{BW}_\rho^\text{GS}(s,m_\rho,\Gamma_\rho)$}.
A fit of this $\rhoz$ shape to the data is only slightly better than the fit of the Breit--Wigner shape given by Eq.~\ref{eq:bw} ($\chindf=290.0/34$, $p$-value $=2\times10^{-42}$). 
A good-quality fit to the data is achieved following the prescription used by the BaBar collaboration
to describe a large $e^+e^-\to\pi^+\pi^-(\gamma)$ data sample~\cite{Lees:2012cj},
\begin{align}
    \M=\frac{\ppi(s)}{\ppi(m_\rho)} \left\{ \mathrm{BW}^\text{GS}(s,m_\rho,\Gamma_\rho)\,\left[1 \right.\right.
    +& \left.\left.A_\omega^\text{GS}\,e^{i\phi_\omega}\,\mathrm{BW}_\omega(s,m_\omega,\Gamma_\omega)\right] \right. \label{eq:gsrhozome}\\
    +& \left. A_{\rho'}^\text{GS}\,e^{i\phi_{\rho'}}\,\mathrm{BW}^\text{GS}(s,m_{\rho'},\Gamma_{\rho'})\right\}. 
    \notag
\end{align}
\begin{figure}[tb]
  \begin{center}
    \includegraphics[width=0.7\linewidth]{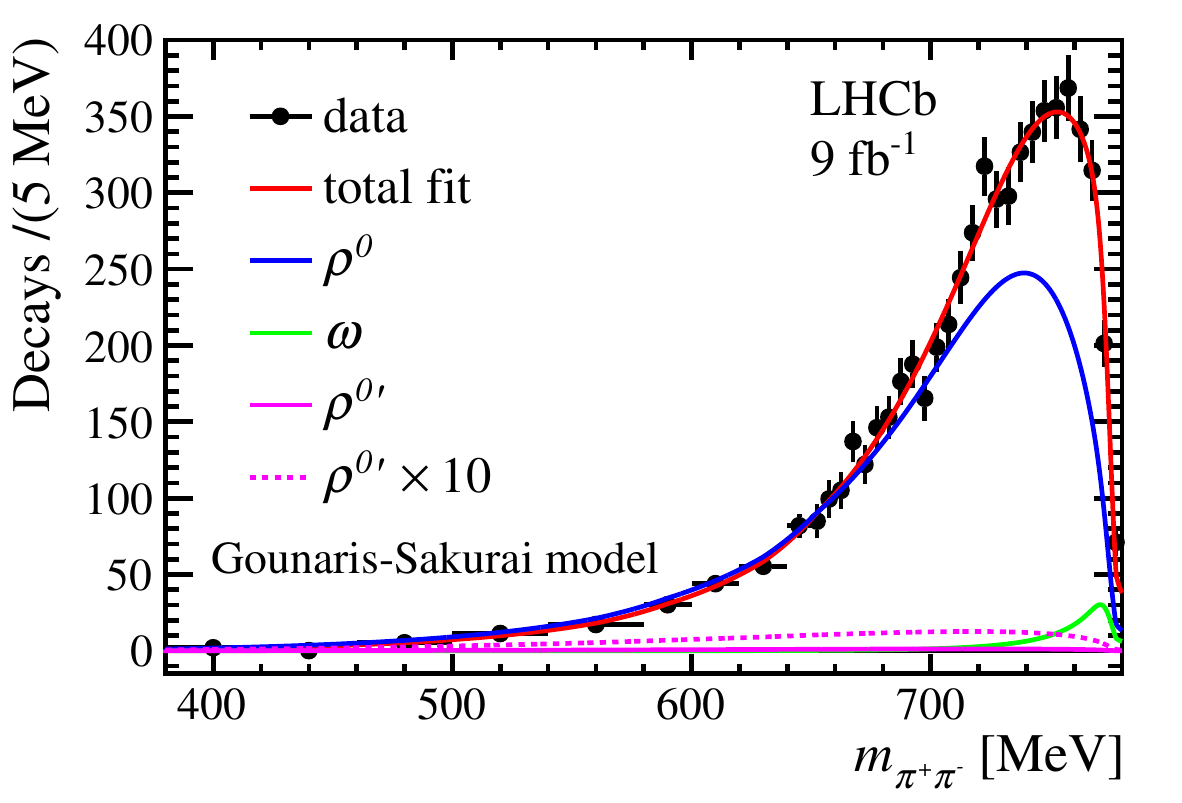}
    \vspace*{-0.5cm}
  \end{center}
  \caption{
    \small 
        Distribution of $\mpp$ in $\theX\to\pip\pim\jpsi$ decays,
        fit with the Gounaris--Sakurai model. The dashed line represents the $\rho'$ contribution multiplied by a factor of 10.
        }
  \label{fig:gs}
\end{figure}
Following this work, a simple Breit--Wigner amplitude for the $\omega$ meson is used,
$\mathrm{BW}_\omega(s,m_\omega,\Gamma_\omega)={m_\omega^2}/{(m_\omega^2-s-i m_\omega \Gamma_\omega)}$,
and the $\rhoz$, $\omega$ and $\rho'$ masses and widths 
are taken from Table VI of Ref.~\cite{Lees:2012cj}. 
The term including the $\rhoz$ and $\omega$ resonances (Eq.~\ref{eq:gsrhozome}) is 
equivalent to Eqs.~\ref{eq:T12}-\ref{eq:alpha2}, and thus originates from the coupled-channel approach. 
No complex phase is expected in this approach, thus $\phi_\omega$ is set to zero. Using the small phase, consistent with zero, obtained by the BaBar collaboration, gives almost identical results.
Since, the $\rho'$ contribution is not added via the $K$-matrix, its phase can be different from zero and is fixed to the central value of the BaBar result, $\phi_{\rho'}=3.76\pm0.10$ rad. 
The $\mpp$ distribution, fit with this model, is shown in Fig.~\ref{fig:gs}.
The fit quality is excellent ($\chindf=24.8/32$, $p$-value $=0.81$),
matching the fit quality of the default fit.
The $\rho'$ significance is $3.1\sigma$ using Wilks' theorem~\cite{Wilks:1938dza}. 
Its production parameter relative to the $\rhoz$ meson, $A_{\rho'}^\text{GS}=0.302\pm0.099$, is consistent within the large uncertainty 
with the value obtained by the fit to the BaBar data, $0.158\pm0.018$~\cite{Lees:2012cj}.
The $\omega$ significance is $7.8\sigma$, and its production parameter, 
$A_\omega^\text{GS}=0.0171\pm0.0024$, is an order of magnitude larger than that obtained by the BaBar collaboration, $0.001644\pm0.00061$~\cite{Lees:2012cj}. 
This is not surprising, since in $e^+e^-$ collisions, the $\rhoz$ meson is produced via electromagnetic interactions, which do not follow isospin symmetry, while in $\theX$ decays, the $\rhoz$ state is produced via strong interactions, which suppress isospin violating decays. 
The $\omega$ fractional contributions are consistent with the default model, $\Rall=0.221\pm0.024$,
$\Rzero=0.021\pm0.005$ and $\Riso=0.028\pm0.007$. 
Here, $\Rall=1-{\cal R}_{\rho+\rho'}$, where 
${\cal R}_{\rho+\rho'}=0.780\pm0.023$ is the coherent fit fraction of the $\rhoz$ and $\rhoz'$ contributions together. The individual fit fractions are 
${\cal R}_{\rho}=0.833\pm0.037$ and ${\cal R}_{\rho'}=0.013\pm0.008$.

\section{Extended $\theX\to\pip\pim\jpsi$ decay phase space}

\begin{figure}[tb]
  \begin{center}
    \includegraphics[width=0.7\linewidth]{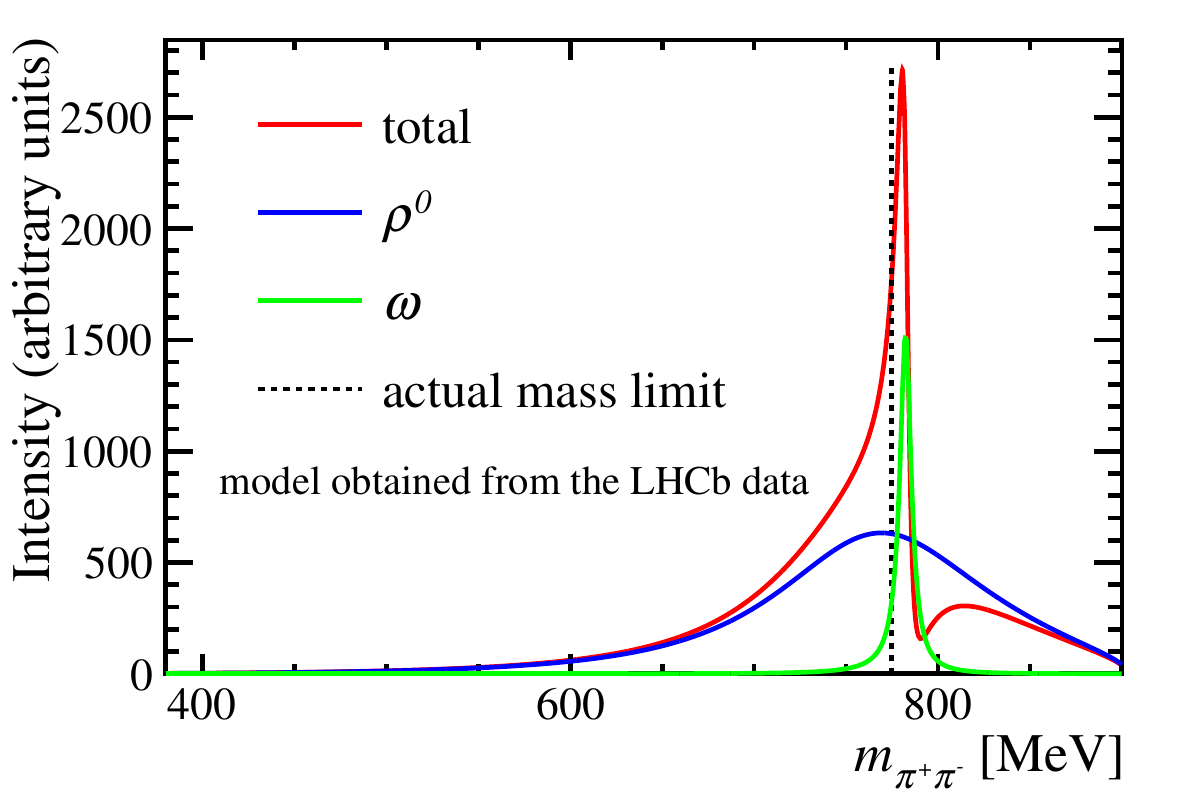}
    \vspace*{-0.5cm}
  \end{center}
  \caption{
    \small 
    Amplitude model for $\theX\to\pip\pim\jpsi$ decays, obtained by the default fit to the LHCb data (Fig.~\ref{fig:default}), with the phase-space integration limit extended by setting the $\theX$ mass to $4000~\mev$. The actual phase-space limit, imposed by the measured $\theX$ mass, is indicated by the vertical dashed line. Neither mass resolution nor detector efficiency are applied here.
        }
  \label{fig:ext}
\end{figure}

To estimate the ratio of the isospin violating to isospin conserving $\theX$ couplings,
the mass of the $\theX$ state in the model obtained by the default fit to the dipion mass distribution is replaced by 
$4000\mev$. 
The extended decay phase space is integrated over both the $\rhoz$ and $\omega$ resonant peaks, as illustrated in Fig.~\ref{fig:ext}.
The obtained ratio, ${g_{\theX\to\rhoz\jpsi}}/{g_{\theX\to\omega\jpsi}}=0.29\pm0.04$, is 
an order of magnitude larger than the value for the well established charmonium state,
${g_{\psi(2S)\to\piz\jpsi}}/{g_{\psi(2S)\to\eta\jpsi}}=0.045\pm0.001$. 

\clearpage


\addcontentsline{toc}{section}{References}
\bibliographystyle{LHCb}
\bibliography{main,standard,LHCb-PAPER,LHCb-CONF,LHCb-DP,LHCb-TDR}

\newpage
\centerline
{\large\bf LHCb collaboration}
\begin
{flushleft}
\small
R.~Aaij$^{32}$\lhcborcid{0000-0003-0533-1952},
A.S.W.~Abdelmotteleb$^{50}$\lhcborcid{0000-0001-7905-0542},
C.~Abellan~Beteta$^{44}$,
F.~Abudin{\'e}n$^{50}$\lhcborcid{0000-0002-6737-3528},
T.~Ackernley$^{54}$\lhcborcid{0000-0002-5951-3498},
B.~Adeva$^{40}$\lhcborcid{0000-0001-9756-3712},
M.~Adinolfi$^{48}$\lhcborcid{0000-0002-1326-1264},
H.~Afsharnia$^{9}$,
C.~Agapopoulou$^{13}$\lhcborcid{0000-0002-2368-0147},
C.A.~Aidala$^{77}$\lhcborcid{0000-0001-9540-4988}
% ...

{\footnotesize \it

% ...
$^{9}$Universit{\'e} Clermont Auvergne, CNRS/IN2P3, LPC, Clermont-Ferrand, France\\
% ...
$^{13}$LPNHE, Sorbonne Universit{\'e}, Paris Diderot Sorbonne Paris Cit{\'e}, CNRS/IN2P3, Paris, France\\
% ...
$^{32}$Nikhef National Institute for Subatomic Physics, Amsterdam, Netherlands\\
% ...
$^{40}$Instituto Galego de F{\'\i}sica de Altas Enerx{\'\i}as (IGFAE), Universidade de Santiago de Compostela, Santiago de Compostela, Spain\\
% ...
$^{44}$Physik-Institut, Universit{\"a}t Z{\"u}rich, Z{\"u}rich, Switzerland\\
% ...
$^{48}$H.H. Wills Physics Laboratory, University of Bristol, Bristol, United Kingdom\\
% ...
$^{50}$Department of Physics, University of Warwick, Coventry, United Kingdom\\
% ...
$^{54}$Oliver Lodge Laboratory, University of Liverpool, Liverpool, United Kingdom\\
% ...
$^{62}$Syracuse University, Syracuse, New York, United States\\
% ...
$^{77}$University of Michigan, Ann Arbor, Michigan, United States, associated to $^{62}$\\
\bigskip
% ...
}
\end{flushleft}

\end{document}